# Synergizing Roadway Infrastructure Investment with Digital Infrastructure for Infrastructure-Based Connected Vehicle Applications: Review of Current Status and Future Directions

Sakib Mahmud Khan[1]*, Mashrur Chowdhury[2], Eric A. Morris[3], Lipika Deka[4]

[1]PhD Candidate, Glenn Department of Civil Engineering, Clemson University, Clemson, SC 29634, USA; email: sakibk@g.clemson.edu

[2]PhD, Professor, Glenn Department of Civil Engineering, Clemson University, Clemson, SC 29634, USA; email: mac@ clemson.edu

[3]PhD, Associate Professor, Department of City Planning and Real Estate Development, Clemson University, Clemson, SC 29634, USA, email: emorri7@clemson.edu

[4]PhD, Lecturer in Computer Science, Faculty of Technology, De Montfort University, Leicester, LE1 9BH, UK, email: Lipika.deka@dmu.ac.uk

*Corresponding author

## ABSTRACT

The safety, mobility, environmental, energy, and economic benefits of transportation systems, which are the focus of recent Connected Vehicles (CVs) programs, are potentially dramatic. However, realization of these benefits largely hinges on the timely integration of the digital technology into the existing transportation infrastructure. CVs must be enabled to broadcast and receive data to and from other CVs (Vehicle-to-Vehicle, or V2V, communication), to and from infrastructure (Vehicle-to-Infrastructure, or V2I, communication) and to and from other road users, such as bicyclists or pedestrians (Vehicle-to-Other road users communication). Further, for V2I-focused applications, the infrastructure and the transportation agencies that manage it must be able to collect, process, distribute, and archive these data quickly, reliably, and securely. This paper focuses V2I applications, and studies current digital roadway infrastructure initiatives. It highlights the importance of including digital infrastructure investment alongside investment in more traditional transportation infrastructure to keep up with the auto industry's push towards connecting vehicles to other vehicles. By studying the current CV testbeds and Smart City initiatives, this paper identifies digital infrastructure components (i.e., communication options and computing infrastructure) being used by public agencies. It also examines public agencies' limited budgeting for digital infrastructure, and finds current expenditure is inadequate for realizing the potential benefits of V2I applications. Finally, the paper presents a set of recommendations, based on a review of current practices and future needs, designed to guide agencies responsible for transportation infrastructure. It stresses the importance of collaboration for establishing national and international platforms for the planning, deployment, and management of digital infrastructure to support connected transportation systems across political jurisdictions.

## INTRODUCTION

Advances in communication technology and data processing capabilities furnish the potential for vehicles to "talk" to each other (via Vehicle-to-Vehicle communication, or V2V), to pedestrians (via Vehicle-to-Pedestrian communication, or V2P) as well as to transportation infrastructure (via Vehicle-to-Infrastructure communication, or V2I). Potential benefits from real-time communication between the elements of the transportation system are dramatic (Chang et al., 2015, He et al., 2012). For example, Connected Vehicles, or CVs (also referred to as "Vehicles with Connectivity"), which broadcast their data to infrastructure and other vehicles, could give drivers advance warning of impending collisions in time to avert dangerous circumstances, dramatically reducing crash damage, injuries, and fatalities. V2I connectivity between vehicles and "digital roadways," which feature roadside devices and backend computation infrastructure, could ensure safe and efficient traffic management in real time, which is not present on public roads today. CVs can benefit the environment with 9,400 tons of annual emission savings for an area covering 45 kilo-meters (28-miles) of US-75 in Dallas, TX. As reported by Chang et al. (2015), about 27% of the delay can be reduced for six intersections in Anthem, AZ, and 11% of the fuel consumption can be eliminated for a 10.5 kilo-meter (6.5 mile) segment of El Camino Real, CA by V2I applications. Further, in the US, roughly 575,000 annual crashes at intersections could be avoided with the use of V2I (Chang et al., 2015). Ultimately, the marriage of Automated Vehicle (AV) technology with advanced communication and data processing technology has the potential to revolutionize auto travel in ways not seen since the introduction of the auto itself (NHTSA 2017, Shladover 2013).

Figure 1 shows the typical roadway digital infrastructure components for a connected vehicular environment (Chowdhury et al. 2017, Lu et al., 2014). Such digital infrastructure is a component of Transportation Cyber-Physical Systems (TCPS). In an environment based on TCPS, CVs will wirelessly communicate with Roadside Units (RSU), which both communicate and process data. Based on the application requirements, additional processing units can be integrated with the RSUs to further increase their data processing capabilities. Such processing units may include commercial computation units such as Intel's Next Unit of Computing (NUC), or ASUS's VivoPC. Data from multiple RSUs will be forwarded to the backend infrastructure, which could be either cloud servers (e.g., Amazon AWS, Microsoft Azure, IBM cloud) or local Traffic Management Center (TMC) servers. These servers would have data storage, processing, and management tools to support CV applications. Data would include real-time, near real-time, and historic data.

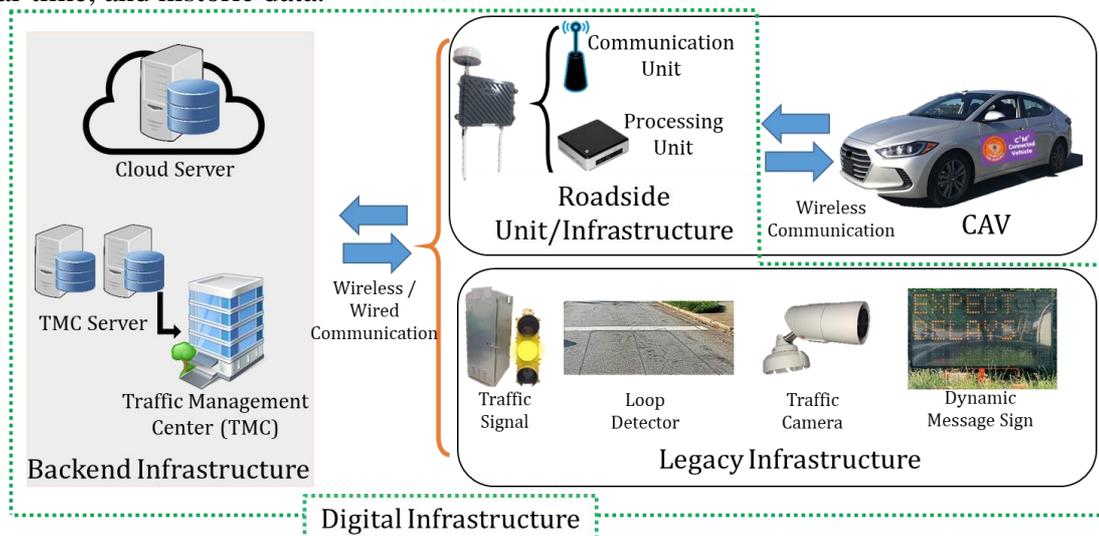

Figure 1. Roadway Digital Infrastructure Components.

Spurred by governments, automakers are rapidly moving toward incorporating communication technology in new vehicles. Communication options such as Long Term Evaluation (LTE) or Wi-Fi already exist in some vehicles. In the US, General Motors has already introduced Dedicated Short-Range Communication (DSRC) technology in its 2017 Cadillac CTS sedans for the purpose of V2V communications; Toyota will include DSRC in Lexus cars from 2021 (Uhlemann, 2018). General Motors also provides the OnStar service, which is an in-vehicle two-way communication system using cellular networks to enhance safety, security and entertainment (He et al, 2017). Similar types of wireless communication services also exist for Ford (SYNC), Volkswagen (Car-Net), and BMW (TeleService). Some brands, such as Audi, Chevrolet, Ford, and Buick, also provide wireless infotainment (i.e., in-vehicle Wi-Fi hotspots based on 4G LTE). In Europe, from April 2018 all new vehicles have the eCall facility to call emergency services in case of crashes (EU 2015).

However, to realize the maximum potential of CVs, public agencies must keep pace with the auto industry. Roadway infrastructure must be upgraded with digital communications infrastructure that evolves with increasing CV penetration levels. This will create an environment suitable for fostering beneficial V2I innovations, such as the V2I safety applications listed by the Connected Vehicle Reference Implementation Architecture (e.g., Curve Speed Warning, Pedestrian in Signalized Crosswalk Warning, Red Light Violation Warning, Warnings about Upcoming Work Zones, etc.) (CVRIA 2018). To benefit from all V2I applications, public agencies need to decide on the type of computing infrastructure (i.e., centralized or distributed) and the communication options (e.g., DSRC, LTE, Wi-Fi) needed to implement a reliable, scalable and connected TCPS. With a centralized computing infrastructure, a TMC server can act as the single computing node/processor to process the CV application locally, whereas a distributed computation infrastructure requires the computation steps to be divided among the different nodes (i.e., RSUs, TMC servers, cloud servers) to minimize computation time and processing costs (Pourebrahimi et al., 2005).

For digital infrastructure investment, proper planning, design, deployment, operations and maintenance are needed. In terms of which phase presents the biggest obstacles, Zmud et al. (2017) found maintenance cost to be the biggest unknown, as this may exceed the initial cost of the deployment of the technology.

Public transportation agencies need to allocate a budget to instrument roadways under their jurisdiction to capture data from CVs, such as traffic volume and speed data. This may not be easy, as expenditure on digital infrastructure must be justified by public agencies which operate in a constrained fiscal environment. In this paper, we discuss the reasons for investment in digital infrastructure for V2I applications, followed by a review of communication options for V2I applications, TCPS computing infrastructures, and existing testbeds. Finally, we highlight current political, technical and investment challenges and future directions so that digital infrastructure deployment will succeed and provide maximum return on investment.

**WHY INVEST IN CONNECTED AND COMPUTERIZED VEHICLES AND ROADWAY INFRASTRUCTURE?**

Even in the absence of vehicle automation, data connectivity in transportation systems promises myriad benefits for travellers and society as a whole. For example, V2I communication will lead to less time-consuming and more ecologically friendly driving. A federal report discussed an integrated eco-corridor management decision support system, which could save 323,000 gallons of fuel annually on a 32 kilo-meter (20-mile) section of I-15 section in San Diego, CA, and 981,000 gallons on an area covering 45 kilo-meter (28-miles) of US-75 in Dallas, TX (Chang et al., 2015)

. In future, these savings will come from utilizing the real-time traffic condition information broadcasted from increasing number of CVs, which will alert drivers, and ultimately their vehicle control systems when such technology exists. Existing studies have found upcoming congestion and incidents can be accurately identified using CV data (Ma et al., 2009, Ma et al., 2012, Khan et al., 2017b). Further benefits from V2I will include better traffic management in work zones (e.g., alerts to motorists to avoid congested routes), better data on infrastructure use for supporting the work of transportation planning and engineering agencies, more timely and accurate condition assessments of transportation infrastructure, and optimized route planning for wireless power transfer for connected electric vehicles. Table 1 provides examples of the potential benefits from receiving data through CV applications for different stakeholders. The applications listed in this table are V2I-based applications, however a few also include V2V and V2P connectivity.

Table 1: TCPS Stakeholders' Benefits

| Stakeholder | Information Received | Application Type | Benefit |
|---|---|---|---|
| **Drivers / CVs** | Information on potential collisions, harsh braking of vehicles in front, hazards at blind corners and intersections, and road obstructions such as construction zones for route planning | V2V and V2I | Automated braking with connected vehicle warning systems: fatality reduction of 37-86 percent in South Australia (simulation study) (ITS Benefits, 2018) |
| | Information on signal phase and timing to maintain an optimized speed through green phases | V2I | Predictive cruise control using traffic signal information: fuel consumption reduction of 24 percent (urban scenario) and 47 percent (suburban scenario) in South Carolina (simulation study) (ITS Benefits, 2018) |
| | Warnings about hazardous material and road conditions such as slippery surfaces, floods, potholes, etc. | V2V and V2I | Data networking and GPS tracking: benefit-cost ratio up to 7.2:1 for HAZMAT trucking in the US (field test) (ITS Benefits, 2018) |
| | Information about points-of-interest such as parking, gas stations, restaurants, etc. | V2V and V2I | On-street parking space information: cruising time reduction by 5-10 percent (simulation study) (ITS Benefits, 2018) |
| **Automated vehicles** | Traffic signal information via V2I communication, vehicle information for Cooperative Adaptive Cruise Control (CACC) | V2V and V2I | 91 percent delay reduction and 82 percent fuel saving for CACC application compared with conventional signal control without CACC (simulation study) (Zohdy 2012) |
| **Vulnerable Road Users (VRUs) such as pedestrians and cyclists** | Early warnings about potential collisions when VRUs approach crosswalks, blind corners or intersections with traffic signals | V2P, V2V and V2I | Vehicle turn warning at crosswalk: 23 percent of 27 pedestrians avoided collision with bus in Portland (field test) (ITS Benefits, 2018) |
| **Traffic Management Centers (TMC)** | Information about current traffic conditions such as traffic flow, congestion, and accidents | V2V and V2I | Getting information from CVs may increase capacity by 273 percent (theoretical modeling and analysis) (ITS Benefits, 2018) |
| | Information about broken down vehicles or incidents | V2I | Incident spot guidance and alerts to approaching connected vehicles and emergency responders: network delay |
| | | | reduction up to 14 percent in CA (simulation study) (ITS Benefits, 2018) |
| | Information about emergency evacuation situations | V2I | Wireless route guidance during evacuation: congestion reduction of 20 percent in Louisiana (simulation study) (ITS Benefits, 2018) |
| **Government and city planners** | Information useful for planning bus and other public transport routes, road capacity improvements, etc. | V2I | More accurate network-level performance measures (compared to Bluetooth sensors and probe vehicles) and vehicle-level travel behavior data (compared to GPS units and mobile phone data) for transportation planning applications (field test) (Deering 2016) |
| | Information to effectively plan new land use developments | V2I | |

For these expected benefits to materialize, public transportation agencies need to accelerate CV application deployment efforts. As shown in Figure 2, according to a US survey conducted in 2016, 59 (out of 95) transportation agencies (including both state and local agencies) have shown interest in deploying CV applications for freeway management, and 95 (out of 274) agencies plan to deploy them for arterial management (ITS Deployment Tracking 2018). It is noteworthy that a relatively higher percentage of agencies are interested in deploying CV applications for freeway management (62%) than on arterials (34%).

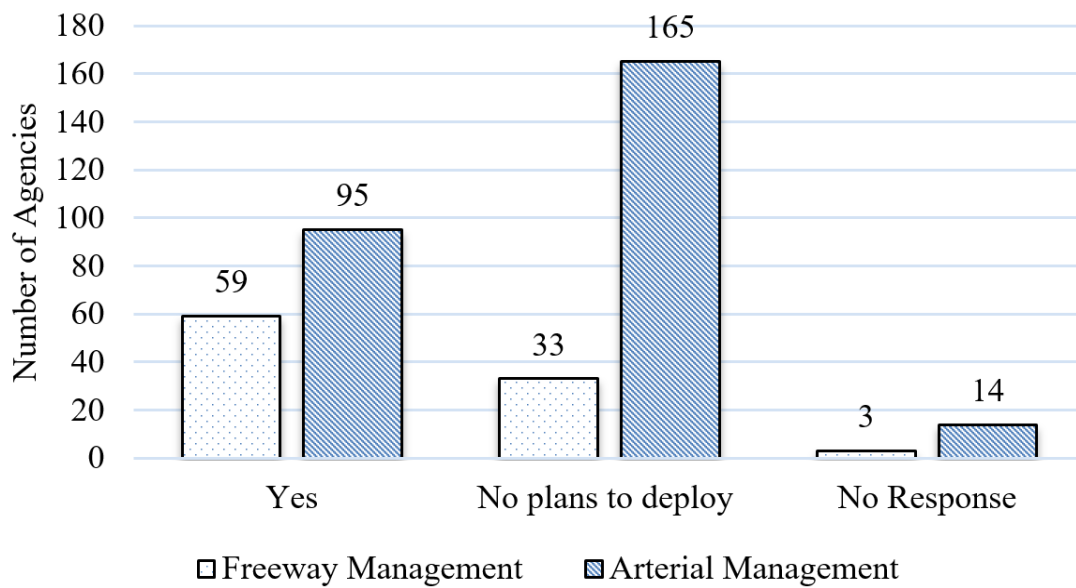

Figure 2. US Agencies Willingness to Deploy CV Applications (Data from ITS Deployment Tracking 2018).

To be sure, there is the potential for problems arising from CVs and AVs. With improved mobility, more people will be attracted to the roads, which will increase vehicle-miles travelled (VMT) (Hörl et al., 2016). Empty vehicles will travel in between drop-offs and pickups for passengers and goods, adding to congestion. AV use may substitute for the use of transit, as the former will have no first mile\last mile issue. The potential impact of AVs and CVs on land use is somewhat ambiguous, as they may encourage dense development in cities or sprawling development in the suburbs (Bagloee et al., 2016). If all connected vehicle drivers get the same navigation advice, or ask for signal priority simultaneously, network efficiency will be adversely affected. Solutions for these problems are possible. For example, in a real-time connected environment, the potential of achieving system equilibrium, with increasing penetration levels of CVs and AVs, will increase, which will reduce the risk of network overloading (Bagloee et al., 2017). Also, intelligent algorithms can address the problem of all CAVs taking the recommended route to avoid congestion that could produce the unintentional consequence of further corridor-level or network-level delay. Dai et al. (2017) observed that the possibility of network-level congestion in their simulated network would be less if only 70% of vehicles could be routed to the first-choice route rather than all vehicles. With proper planning, CVs and AVs can be synergistically integrated with the non-connected vehicle stream in a sustainable fashion so that adverse impacts of CVs and AVs could be mitigated. Krechmer et al. (2015) has discussed how CVs can be integrated in the planning process by both state and local transportation agencies in a coordinated way so that CVs can provide positive benefits.

It must be stressed here that V2I connectivity is an enabler which will allow AVs to reach their full potential (Litman 2017). Although AVs will likely need to be able to operate without connectivity in order to be not dependent on external infrastructure for safe operations in case connectivity is not available, connectivity would dramatically improve AVs' functioning. The United States Department of Transportation (USDOT) has highlighted the importance of connectivity between AVs and other vehicles and infrastructure in the Automated Vehicle Research webpage (ITSJPO 2018). Duran et al. (2013) used fault tree analysis to study the risks inherent in the use of AV sensors (LIDAR and camera), and found that sensor failure would be the leading cause of pedestrian fatalities. Based on 2016 AV testing data provided by California Department of Motor Vehicles, AV sensor-related hardware and software failure caused five to eighteen percent of the total incidents that occurred during AV field testing (Bhavsar et al., 2017). Further, connectivity will lower failure rates of AVs by providing additional data beyond the coverage area of the AV sensors (e.g., LIDAR, camera). For example, tightly packed platoons of vehicles operating at high speeds will be possible in an AV environment, but the considerable safety challenges posed by this strategy would be dramatically reduced if vehicles at the front of the platoon communicated their speed and position to followers in real time. By providing information about current and planned actions of leading connected AVs, connectivity will help follower AVs to take early and appropriate responses. In addition, intelligent intersections will feature signals which change phases based on current and future traffic conditions as determined by communications from oncoming CVs; this ultimately may even dispense with the need for traditional signals altogether (Fayazi and Vahidi, 2017), as vehicles can be woven through the intersection, dramatically increasing throughput. Also, external connectivity, including connectivity with roadside or roadway infrastructure, will reduce the extent and cost of the sensors and computing systems AVs may be required to carry on-board. Shladover (2013) found that connectivity can augment data captured by AV sensors, which will: (a) reduce the impact of AV sensor uncertainty, (b) limit processing lags for filtering AV sensor data, and (c) capture information beyond AV sensors' coverage. To truly maximize the benefits of AVs, then, connectivity will be essential.

**RESEARCH ON CV COMMUNICATION AND COMPUTING INFRASTRUCTURE**
*Communication Options for V2I Applications*

For V2I applications, digital infrastructure consists of embedded sensors and backend computation infrastructure, which can exchange real-time data between road management agencies or other data providers and users via a reliable communication network. Road management agencies can use any one of, or combinations of, several communication options: Dedicated Short-Range Communication (DSRC), Cellular technologies (such as 4G, 5G), Wireless Fidelity (Wi-Fi), Worldwide Interoperability for Microwave Access (WiMAX), Bluetooth, etc. Latency, bandwidth, cost, communication range, and the reliability of different communication options vary for different applications. In 2016, the ITS Joint Program Office conducted a survey on US public agencies' communication technology adoption, and received responses from 272 arterial management and 99 freeway management agencies (ITS Deployment Tracking 2018, USDOT 2018). Figure 3 shows the options adopted by these agencies to enable communication between multiple ITS devices, or between ITS roadside devices and a central processing location. In general, cellular LTE is the most widely adopted wireless communication option. The other wireless options that have been adopted include Wi-Fi, WiMAX, DSRC, and microwave.

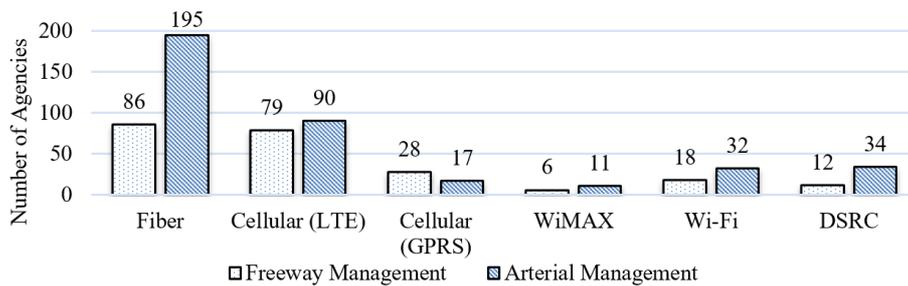

Figure 3. Communication Options Deployed by US Agencies (Data from ITS Deployment Tracking 2018).

Fifty-three of the 78 cities participating in the UDSOT Smart City Challenge proposed implementing DSRC connectivity to enable communications between vehicles and infrastructure (USDOT, 2016). On the other hand, downtown Kansas City features an intelligent Wi-Fi network (KCMO 2016; Boissevain, 2018) which wirelessly connects and adjusts smart street lights based on pedestrian presence. Such wireless connectivity could provide monetary and environmental benefits once CVs are deployed (Chang et al., 2015). Another option would be fifth-generation (5G) cellular communications systems, which feature greater range (up to 32 kilo-meters (20 miles)) and increased throughput compared with DSRC (Cordero 2016). The cellular alternative to IEEE802.11p/DSRC is being heavily backed by vehicle manufacturers and network operators as is evident from establishment of the 5G Automotive Association (5GAA), which was set up in 2016 (5GAA, 2018).

Table 2: Characteristics of Wireless Communication Networks

| Communication Options | Single-hop Latency | Range | Spectrum |
|---|---|---|---|
| **DSRC** | 0.0002 sec | 300 meter (1000 ft.) | 5.85-5.925 GHz (US) 5.875-5.925 GHz (Europe) 5.77-5.85 GHz (Japan) |
| **Cellular LTE 4G** | 0.01 – 0.02 sec | < 29 kilo-meters (18 miles) | Different ranges between 450 MHz to 3.6 GHz |
| **Cellular LTE 5G** | 0.0001 sec | 32 kilo-meters (20 miles) | Different ranges between 600 MHz amd 100 GHz |
| **Wi-Fi (802.11)** | 0.006 sec (for 2.4 GHz) 0.0009 sec (for 5 GHz) | 31 meters (100 ft.) | 2.4 and 5 GHz |
| **WIMAX** | < 0.01 sec | 50 kilo-meters or 31 miles (with line of sight) | 2.3, 2.5, 5.8 GHz (US) 2.5, 3.5, 5.8 GHz (Europe) |

|  |  | 5.5-10 kilo-meters or 3.5-6 miles (with no line of sight) |  |
|---|---|---|---|

Table 2 shows that different wireless communication options have different single-hop latency, range, and allocated spectrum (Chintapalli et al. 2013, de Carvalho et al. 2017, Ghadialy 2015, Hpbn 2018, Lee et al. 2018, Odiaga et al. 2016, Remy and Letamendia 2014, Shabbir and Kasif 2018, Zhou et al. 2009). Because of the different strengths and weaknesses of the different technologies, it is recommended that a real-world connected TCPS make use of a heterogeneous network which can support multiple applications at the same time (Siegel et al. 2018). A Heterogeneous Wireless Communication Network (HetNet) permits selection of wireless network options (e.g., Wi-Fi, LTE and DSRC) to exchange data between data users and data providers based on communication delay, availability of communication options, communication coverage area, and communication reliability, considering the temporal and spatial requirements of the V2I applications. Prior research has studied the applicability of heterogeneous networking for TCPS (Chowdhury et al., 2017, Dey et al. 2016), which can be leveraged by public agencies.

*TCPS Computation Options for V2I Applications*

On-premise computing (e.g, Traffic Management Center (TMC) servers), cloud computing, and edge/fog computing are available for V2I applications. Robust and reliable algorithms for V2I applications often have to meet real-time and/or near-real time processing requirements (Zheng et al. 2015). To make the TCPS scalable and resilient with increasing numbers of CVs, edge computing is a viable option for public agencies. An "edge" is any computing resource (e.g., an On-Board Unit (OBU), RSU, server) which can help with data storing, processing, and service request distribution along the path between CV data sources and CV data consumers (Shi et al., 2016). Edge computing paradigm in a CV environment can be defined as computational services to run CV applications in computing devices, such as in RSUs, that are distributed by nature and close to the data sources (e.g., CVs), which facilitates low data loss and data communication latency between CV the data sources and computational services. By distributing the computation to different edges, edge computing also ensures high bandwidth. As discussed by Chowdhury et al. (2017) on their work on the Clemson University CV Testbed (CU-CVT), mobile entities such as CVs and pedestrians with connected wearable devices, RSUs, and a backend server are different edges. These edges have different levels of computation capabilities and memory storage to support multiple CV application requirements with increasing CV penetration. A white paper by the 5GAA (2017) demonstrates a number of diverse cases where edge computing will be particularly effective.

**CURRENT DIGITAL INFRASTRUCTURE INITIATIVES IN CV**
*CV Testbeds and Initiatives*

V2I applications will require common standards to insure interoperability, whether across different makes of vehicles or across infrastructure in different political jurisdictions. Some progress has been made on this. In the US, pilot digital infrastructure initiatives have been mostly led by state agencies and academic institutions with industry collaboration. The Mcity initiative, a testbed for V2V and V2I development run under the aegis of the University of Michigan in Ann Arbor, is an example of government, academia and industry collaboration; it enables CV testing to be done in a safe and controlled environment before deployment in a public environment (M-city, 2017). The USDOT has also supported research on centralized digital infrastructure by sponsoring several CV research projects. In a program funded by the USDOT, New York, Florida and Wyoming have been selected as CV pilot sites in year 2015. The instrumentation descriptions for these and other CV deployment sites are shown in Table 3 (Cregger et al., 2012, Dickey et al. 2010, Misener and Shladover 2006). These pilot sites' instrumentation requirements include deploying RSUs, connecting RSUs with back-end computational infrastructure via wired/wireless communication networks, and developing computing infrastructure to store and process the data.

Table 3: CV Deployments in the US

| State | Project Name | State DOT Role | Deployment Site | Deployment Description | Transportation Applications | Communication Options Used | Computing Infra-structure |
|---|---|---|---|---|---|---|---|
| **Virginia** | Virginia Connected Corridors | Partnered with Virginia Tech Transportation Institute and Virginia Transportation Research Council | Smart Road near Blackburg, VA: 3.5 kilo-meters (2.2 mile) two-lane road from Transportation Research Drive to Wilson Creek Bridge | 10 RSUs | Weather impact on transportation, OEM-desired applications | DSRC, Fiber Optic | Cloud-based centralized system |
| | | | Fairfax in Northern Virginia, including sections of Interstate 66, Interstate 495, U.S. 29, and U.S. 50 | 66 RSUs, 50 highly instrumented light vehicles, five District Department of Transportation (DDOT) safety service patrol (SSP) trucks | OEM-desired applications, road surface condition data | DSRC, LTE | |
| **California** | California Connected Vehicle Test Bed | Owner | 16 kilo-meters (10 mile) segments of 2 routes (in Palo Alto and near the San Francisco Airport), encompassing US 101 and State Route 82 | 9 RSUs along SR-82 | Intersection safety applications, intelligent on-ramp metering, travel time data to vehicles, work zone safety warnings, taking curves over speed warnings | DSRC, LTE | Centralized data management system |
| **Colorado** | E-470 Toll Plaza | Not involved (tested furnished by Kapsch) | 3 lanes next to an existing E-470 highway toll collection system in Aurora, CO | RSUs, 27 instrumented vehicles, cameras, laser units | Road tolling and enforcement. | DSRC | No information available |
| **New York** | NYC CV pilot deployment | Owner | Manhattan arterials (within 14th and 67th street), | 320 RSUs, 10,000 vehicles with after- | Collision warning, blind spot warning, curve speed compliance, | DSRC, cellular (NYCWIN), fiber optic | TMC-based centralized system |

| State | Project Name | State DOT Role | Deployment Site | Deployment Description | Transportation Applications | Communication Options Used | Computing Infra-structure |
|---|---|---|---|---|---|---|---|
| | | | Brooklyn Flashback Avenue, Manhattan FDR freeway | market safety devices | pedestrian in signalized intersection warning | | |
| Florida | Florida Test Bed | Owner | Corridor in Orlando along I-4 and along John Young Parkway/ International Drive/Universal Boulevard | 11 RSUs along I-4 and 16 RSUs at other locations, 41 vehicles | Traffic management | DSRC, Fiber Optic | TMC-based centralized system |
| | Tampa CV Pilot Deployment | Partnered with Tampa Hillsborough Expressway Authority (THEA), USDOT, City of Tampa, etc. | Downtown Tampa | 46 RSUs, 1600 private cars, 10 buses, 10 street cars, 500 pedestrians | Traffic backup warning, wrong-way warning, transit signal priority, traffic flow optimization | DSRC, Wi-Fi, LTE | TMC-based centralized system |
| | Osceola County Connected Vehicle Deployment | Partnered with Osceola County and FHWA | Osceola Pkwy. and Orange Blossom Trail intersection, and Orange Blossom Trail and Poinciana intersection in Kissimmee | 2 RSUs (with capability to run signal phase and timing applications) | Showing Signal Phasing and Timing (SPaT) information on OBUs | DSRC, Fiber Optic | TMC-based centralized system |
| Michigan | Ann Arbor Connected Vehicle Test Site | USDOT partnered with University of Michigan Transportation Research Institute | 117 kilo-meters (73 miles) of roadway in the northwestern part of Ann Arbor | 29 RSUs, 2836 vehicles | Safety benefits of connected vehicles | DSRC, LTE, Fiber Optic | No information available |
| | Southeast Michigan Connected Vehicle Testbed | USDOT-sponsored | Sections of I-96, I-94 (Ann Arbor-metro Detroit), and U.S. 23 (Ann Arbor-Brighton) | 50 RSUs, 9 vehicles | Signal phasing and timing, security credential management system | DSRC, Fiber Optic | Situation data processing center-based centralized system |

| State | Project Name | State DOT Role | Deployment Site | Deployment Description | Transportation Applications | Communication Options Used | Computing Infrastructure |
|---|---|---|---|---|---|---|---|
| **Wyoming** | Wyoming CV Pilot Deployment | Owner | I-80 corridor | 75 RSUs, 400 vehicles | Forward collision warning, work zone warning, spot weather impact warning | DSRC, WyoLink Radio Network, LTE, Wi-Fi | TMC-based centralized system |
| **Arizona** | Arizona Connected Vehicle Test Bed | Partnered with Maricopa County, DOT, University of Arizona | 3.7 kilo-meters (2.3 miles) on arterial, 11 signalized intersections, six freeway interchanges, and 10 other freeway locations in Anthem, AZ | 12 RSUs, 2 MCDOT REACT vehicles, 10 vehicles | Traffic signal control priority for electric vehicles and transit, traffic signal priority applications | DSRC, Wi-Fi, Bluetooth, Fiber Optic | TMC-based distributed system |
| **Minnesota** | Minnesota Connected Vehicle (CV) Pilot Deployment | Owner | I-35W southwest of Minneapolis | 6 RSUs, 600+ vehicles | Maintenance activities | LTE, DSRC | TMC-based centralized system |
| **Pennsylvania** | CMU Cranberry Township and Pittsburgh Test Bed | Partnered with Carnegie Mellon University (CMU), Cranberry Township, and the City of Pittsburgh | 2.9 kilo-meters (1.8 miles) stretch along Route 19 corridor | 11 RSUs | Traffic signal-related applications | DSRC | No information available |
| | | | Baum Boulevard (state route) and Centre Avenue (city road) corridors | 24 RSUs | | | No information available |
| | PennDOT Ross Township Test Bed | Partnered with FHWA | Along McKnight Road (SR 4003) from I-279 to Perrymont Rd/Babcock Blvd in Pittsburgh, PA | 11 RSUs | Traffic signal related applications | DSRC | No information available |
| **South Carolina** | Center for Connected Multimodal Mobility Testbed | USDOT sponsored | 3.2 kilo-meters (2 miles) long stretch along Perimeter Road, Clemson SC | 3 RSUs, 20 vehicles | Queue warning, speed harmonization, heterogeneous network testing | DSRC, Wi-Fi, LTE, Fiber Optic | TMC-based distributed system |

| State | Project Name | State DOT Role | Deployment Site | Deployment Description | Transportation Applications | Communication Options Used | Computing Infra-structure |
|---|---|---|---|---|---|---|---|
| **Utah** | UDOT Redwood Road DSRC Corridor | Owner | Redwood Road (1700 West) from 400 South Street (Salt Lake City) to 8020 South Street (West Jordan) | 30 RSUs, 4 buses | Transit signal priority | DSRC, Fiber Optic | TMC-based distributed system |

In addition to funding several pilot projects, the USDOT has provided guidance through its Connected Vehicle Reference Implementation Architecture program (USDOT, 2017). This initiative aims to support standards development for data collection and communication networks. It has shown how different transportation components, such as vehicles, roadway infrastructure, and data storage and processing infrastructure should exchange data and what types of data should be exchanged. This program serves as an important roadmap towards the future.

Outside the US, Asian and European countries are also active in conducting research and deploying pilot projects involving CVs (Khan et al., 2017a). In Europe, there has been an accelerating effort to deploy CV technologies and make the roads ready for connected vehicles. Table 4 outlines some existing CV deployment sites outside the US. Among other initiatives, the European Commission, through its Europe on the Move strategy, has recently completed its agenda on safe mobility (using mandatory advanced driving features and smarter roads to move toward a goal of zero road fatalities by 2050), clean mobility (with new $CO_2$ emission standards for heavy-duty trucks aiming at a 30% reduction in emissions by 2030), and connected and automated mobility. Four hundred and fifty million euros are being invested to achieve these goals (European Commission, 2018). There is a current investment program in the UK that is allocating £11 billion between 2015 and 2021 for the creation and upgrading of "smart motorways" (Highways England, 2017). These motorways will automatically keep track of congestion to dynamically change speed limits, as well as open hard shoulders as traffic lanes to mitigate congested conditions. While this investment is commendable, and has been successful in terms of reducing congestion, a more forward-looking investment would include CV technologies. A comprehensive detail of UK's 2018 projects can be found at CCAV (2018).

Table 4: CV Deployments outside US

| Country | Deployment Site | Deployment Description | Transportation Applications | Year | Source |
|---|---|---|---|---|---|
| **Multiple countries from European Union (EU)** | 7 intersection sites in 7 participating countries | 150 RSUs, 662 vehicles | Traffic signal violation warning, roadway hazard warning, and intersection energy efficiency improvement | 2013-2015 | Compass4D 2017 |
| **Multiple countries from EU** | Application deployed in intersections and within emergency vehicles, other participating vehicles, and road work sites | Cellular based 3G-4G/LTE mobile communication networks (for C-Roads Belgium) | Emergency vehicle approaching, road works warning, in-vehicle speed limit, intersection safety, weather conditions, and in-vehicle signage | Ongoing | CRoads 2017 |
| **France** | Ile-de-France, Paris-Strasbourg highway, Isère, the ring road of Bordeaux, Bretagne | 3000 vehicles, RSUs at 5 sites (almost 2011 kilometers (1250 miles) of road) | Slippery road warning, road work information | Ongoing | Scoop 2017 |
| **Austria** | Almost 45 kilometers (28 miles) long corridor close to the motorway junctions A2/A23-A4-S1 in Vienna; belongs to the Telematics Consortium | 46 roadside communication points, including 10 traffic lights | Traffic safety and traffic management | 2013 | ECoAt 2017 |

| United Kingdom | Two sites in the UK including 90 kilo-meters (56 miles) of roads | 5G emulation, data storage, controlled to semi-controlled urban environment along 90 kilo-meters (56 miles) of roads | All aspects of real-world CV operation including Mobility-as-a-Service and social impacts | 2019 | Millborrk. 2018 |
|---|---|---|---|---|---|

To motivate public and private stakeholders to adopt common standards which would allow for interoperability, the European Commission has created a common platform to facilitate deployment of CV technologies called the Cooperative Intelligent Transport Systems (C-ITS) initiative. This program, initiated in November 2014, brings together key public and private stakeholders (e.g., government authorities, auto manufacturers, suppliers, service providers, and telecommunication companies) to adopt a common vision and accelerate innovation and deployment of CV technology. Created in consultation with various stakeholders, the C-ITS platform addresses technical, legal and policy issues, such as the underlying communication medium, security and certification, the data integration platform, privacy and liability issues arising from data sharing and usage, standardization, interoperability among stakeholders (particularly across political borders), effective business models, and more. Although the C-ITS platform has been developed, substantial investment and development are required for both vehicles and infrastructure before many socio-economic benefits can be reaped; it has recommended that governments must continue to invest in V2I technology deployment so that private companies see clear benefits and continue investing in in-vehicle technology. C-ITS stresses that since CV technology is currently ready for adoption, and since vehicle manufacturers aim to deploy CV-enabled vehicles in the EU by 2019, setting up the technical and legal infrastructure is urgent (IEEE 2015).

While industry, academia, and governments in the US and Europe aim to be at the forefront of the advancement of the CV technology, there is a marked difference in their approaches. In the US, policy makers have left it to industry to decide which connectivity option to adopt for CVs, and the majority of the automakers and network operators have recently put their support behind cellular technology for V2I communication (terming it Cellular Vehicle-to-Everything, or V2X). This gained traction with the formation of the 5G Automotive Association (5GAA) (5GAA 2018) in 2016. On the other hand, across the Atlantic, the approach is more prescriptive and standards tend to be set more by government, though stakeholders influence policy. The European Commission plans to leverage both Intelligent Transport Systems-G5 (ITS-G5), a wireless communication option similar to DSRC, and cellular communications for vehicular connectivity (5G America, 2018). Based on these communication options, the European Commission plans to finalize the legal framework soon for the implementation of cooperative intelligent transportation systems by 2019.

*Connected Transportation and Smart Cities*

Connectivity in transportation is a key element of the "Smart Cities" concept, or, as a recent report from the USDOT (2014) calls it, the "Smart/Connected Cities" concept. USDOT foresees Smart/Connected Cities as interconnected networks of systems including employment, transportation, public services, buildings, energy distribution, and more. These are referred to as systems of systems, which are linked together by Information and Communications Technologies (ICT). Within the Smart/Connected City, ICTs broadcast and process data about different activities. The Smart/Connected City will use intelligent infrastructure that translates the state of the physical world into data through devices that sense their environment, and collect, exchange, and analyze that data through advances in ICT such as crowdsourcing, Big Data analysis, and gamification.

Big Data is characterized by volume (i.e., large data size which cannot be analysed with traditional data analysis software), velocity (i.e., data coming in real-time, or a certain interval), veracity (i.e., trustworthiness of the data), variety (i.e., data having different formats, and types), and value (i.e., worth or efficacy of the data) (Khan et al., 2017a). In addition to collecting and transmitting data, infrastructure will sometimes receive instructions for action. The goal is to create synergies between smart and programmable infrastructure systems, such as the electricity grid, waste disposal, water distribution, healthcare, and more. Connected transportation is a major element of Smart Cities, as is shown in Figure 4. The digital infrastructure may process information not only on traffic flows and road conditions, but also on related systems such as energy systems (e.g., fuel consumption by vehicles), the environment (e.g., vehicle emissions, hazardous material exposure), and the community (e.g., traveller information, traveller satisfaction).

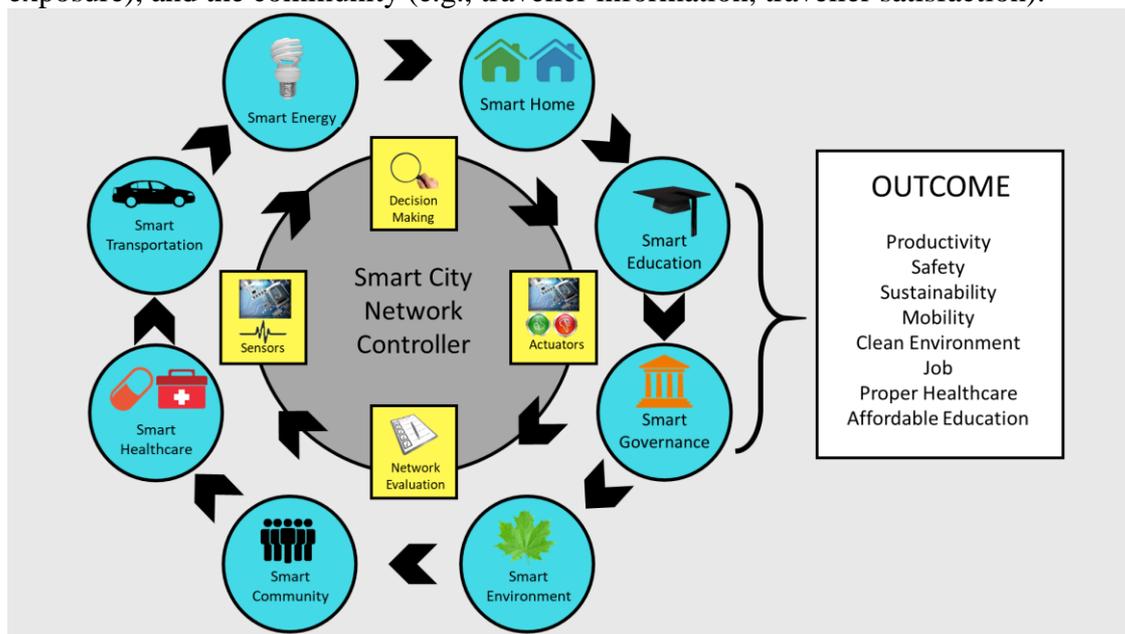

Figure 4. Overview of Smart City Components.

The USDOT has laid out a research agenda to make the Smart/Connected City—particularly its transportation element—a reality. This includes not only developing a connected transportation system, but exploring how this system will interface with other aspects of a Smart City such as the Smart Grid, Smart Homes, etc.; how the system can be used to influence traveller behaviour while ensuring sustainability and a reduced carbon footprint; what role the Internet and mobile devices can play in a Smart/Connected City system; what actors (such as, travellers, private and public agencies) must be engaged to make the Smart/Connected City a reality; and what the social, political, environmental, and economic benefits of a Smart/Connected City may be.

Innovative initiatives around the globe are currently making the Smart/Connected City, with a strong transportation element, a reality. In 2016, USDOT held a Smart City challenge, and selected Columbus, OH, as the winner. The Smart City plan for the City of Columbus the use a number of new technologies, such as connected infrastructure, electric vehicle charging infrastructure, an integrated data platform, and autonomous vehicles to meet the current and future challenges in different areas including transportation, residential and commercial. The European Innovation Partnership on Smart Cities and Communities (EIP-SCC) (European Commission, 2015) brought together cities, residents and industries to develop a number of solutions that have found their way into commercially viable products, start-ups and services, such as the SuperHub tailor-made mobility solution, which suggests to people the most eco-friendly mobility option (European Commission, 2017). In the United Kingdom, a multi-million pound Smart/Connected City project has been launched in the city of Bristol to upgrade the existing infrastructure with the latest sensors and connectivity technology, turning the city into a live laboratory to test and deploy solutions for combating air pollution and traffic congestion while helping to assist the elderly and support the city's trial of AVs (Bristol, 2017). Rio de Janeiro has the world's largest "smart" operations center, created in collaboration with IBM. It collects and analyzes data from myriad sources to optimize city services. The initial focus was on disaster prediction and response, but the program has grown to include transportation, with data being drawn from traffic and transit navigation apps. Another example of a Smart City is Songdo, South Korea, where CISCO, a private company, demonstrated the connected community concept by connecting offices, residences and other buildings (Angelidou, 2014). With the help of remote control systems, residents have the capability to control different functionalities in their homes. This compact city also has an accessible transportation system with widespread provision of public transit, biking, and walking facilities, and the whole city is under surveillance for real-time traffic management. In Amsterdam, an open data platform has been developed with the help of public agencies, utility companies, and other data providers to visualize the energy consumption of the local residents (van den Buuse et al., 2018; Loibl et al., 2014). The open platform provides data for decision-making regarding energy management, and is used by the local agencies. Perhaps the world's leading Smart City is Singapore, where a highly developed data-gathering and analysis system, including sensors, cameras, and GPS devices provides information on traffic and congestion to aid navigation, transit operations, and the congestion tolling program. (For more on these cities, see USDOT (2014).)

## INVESTMENT TRENDS IN DIGITAL INFRASTRUCTURE BY PUBLIC TRANSPORTATION AGENCIES

However, notwithstanding all of these initiatives, public agencies' limited investment in digital infrastructure is clear when examining agencies' ITS deployment and improvement funding. For example, for the fiscal year 2019 the total operating budget requested for Arizona DOT is $33 million more than 2018, yet for statewide Intelligent Transportation Systems (ITS) upgrades and maintenance the requested budget increment is only $2 million (only 0.6% of the total highway budget increment) (Arizona 2017). This additional funding is requested mainly for replacing and updating statewide Closed Circuit TV (CCTV) cameras, Dynamic Message Signs (DMS), and Road Weather Information Systems (RWIS). Arizona DOT is currently under-funded by $500,000 per year for the statewide ITS infrastructure (Arizona 2017). For the Wisconsin DOT, the total proposed allocation for its ITS program, according to the 2015 biennium funding request, is only 0.67% of the total state transportation funding request (Wisconsin 2015). Although the funding request is higher than in prior years, it is not sufficient to implement ITS infrastructure on selected corridors, which includes the installation, replacement and rehabilitation of traffic signals, CCTV, DMS, ramp meters, and related communication networks. According to the State Transportation Improvement Program for the Massachusetts DOT, the proportion of the budget devoted to ITS is very small, at only 1.61 percent (MassDOT 2017). For ITS programs, the budget includes the cost of traffic sensors, CCTV, and DMSs. For the Colorado DOT (CDOT), ITS devices include CCTV, radar devices, RWIS, travel time readers, ramp meters, and automated traffic recorders. According to the proposed budget plan, 3% (i.e., $37 million) of the total DOT budget was to be allocated to be used for ITS programs in 2016. However, the actual budget spent was only $27 million (CDOT 2015). As these data suggest, the ITS program budget in many US states is not sufficient to implement the widespread digital infrastructure for the V2I applications of the future.

In the UK, while investment in ITS infrastructure has recently increased (since 2015 there has been over £1.5 billion in ITS investment in England for upgrading the motorways), much of it is limited to the trunk roads and motorways, primarily in "traditional ITS" technology such as CCTV, DMS and traffic control centres (Trans Scot 2017). In addition, there have been a number of small pots of investment under the C-ITS project funded by the Department of Transport.

# FUTURE DIRECTIONS TO OVERCOME CHALLENGES IMPEDING V2I DEPLOYMENT

## Political Challenges and Opportunities

This lack of resources for V2I comes in the context of increasingly constrained funds for transportation in the US. Although the United States was a world pioneer in terms of funding and building a massive highway system, in recent decades a lack of funding and political will has precluded dramatic new investment in highway infrastructure. For some time, observers have noted that the condition of US roadway infrastructure is suboptimal and declining. The American Society of Civil Engineers issues an annual "Report Card" on the state of roads and bridges in the US: currently, the road system receives a grade of "D" (i.e., poor-fair condition with many roadways approaching the end of their service lives), while bridge infrastructure receives a "C+" (i.e., fair-good condition with many bridges exhibiting signs of general deterioration) (ASCE, 2017). The 2016-2017 *Global Competitiveness Index*, available from the World Economic Forum, ranks the US only thirteenth for overall road quality (World Economic Forum, 2017). Several national-level assessments have called for immediate action and fresh investment in order to repair and upgrade American transportation infrastructure (Pisarski and Reno, 2015, Business Roundtable, 2015). A severe infrastructure and transit funding shortfall in US has thus been identified, worth $846 billion for the seven year (2013-2020) planning timeframe (Zmud 2017).

Clearly, if resources are lacking even to keep pavement in good condition, questions abound about funding for digital infrastructure. This is unfortunate because the technology exists, and has been proven feasible as well as highly beneficial, to integrate travellers, such as drivers, cyclists, and pedestrians, and infrastructure, such as traffic lights, open-road tolling facilities, DMS boards, highway onramp meters, and regional traffic control centers. Thus far, despite promising pilot programs, the political will to deploy this technology on a large scale has proven elusive. This lack of willingness is evident from the funding shortfall and discrepancy between automotive R&D and public investment in traffic management infrastructure. In 2015, global automotive R&D expenditure for 92 auto companies was $109 billion (PWC 2015), while the global traffic management system market was only $4.12 billion (Market Research, 2016). Such an investment mismatch may lead to an environment where the benefits of a smart, connected ecosystem will never fully reach fruition.

In addition to a lack of financial resources, several other hurdles must be surmounted in order to proceed from the research phase to the actual deployment of V2I technologies by the public sector. According to the United Nations Economic Commission for Europe, the impediments inhibiting ITS deployment include a lack of political will, a lack of harmonized ITS deployment policies, and a lack of coordination between public agencies and the private sector (UNECE 2012). Based on prior experience, different agencies/stakeholders will have different perspectives on who should take the lead in investing in, operating, and maintaining digital connected infrastructure. Identifying the roles and responsibilities of stakeholders (as shown in Table 5), and developing a consensus regarding the investment, deployment, operations and maintenance of the digital infrastructure, are the most critical steps in mainstreaming CV technology.

Table 5: Agency Responsibilities for Digital Infrastructure Deployments

| Tasks | Federal Transportation Agency | Regional Transportation Agency | Private Actors |
|---|---|---|---|
| **Standards development** | X | | X |
| **Device specification development** | | | X |
| **Device interoperability checking** | X | | X |
| **Security and privacy control** | X | | X |
| **Communication mandate** | X | | |
| **Device and software manufacture** | | | X |
| **Digital infrastructure deployment and management** | | X | |
| **Outreach program and staff training** | X | X | |

Finally, the legal implications of the digital infrastructure must be addressed at the political level. Responsibilities of various stakeholders must be assigned; for example, legal liability for safety events and security and privacy breaches must be sorted out. In their report published for the European Commission, Brizzolara and Flament (2017) also identified liability, standardization, and government roles as challenges facing vehicle and roadway automation.

*Technical Challenges and Opportunities*

Although the technology has come a long way, some of the technical hurdles still facing V2I remain somewhat daunting. Once CV technology is deployed, the sheer number of vehicles and the extent of transportation infrastructure will generate a tremendous volume of data, and existing facilities come nowhere close to having the capacity to communicate, store and analyze such an unprecedented amount of information. To illustrate, California's freeway Performance Monitoring System (PeMS) is a repository for all fixed-location loop detector data in the state. From the more than 23,000 loops statewide, 2 GB of data are collected by PeMS per day (Choe et al. 2012). In contrast, a single CV can produce 25 GB of data per hour (Hitachi 2015), and AVs can produce more than 165 GB per hour (Velde et al., 2017). This enormous disconnect calls for dramatic investment to upgrade existing information technology infrastructure to make it fully capable of handling Big Data from both vehicles and infrastructure. National, state, regional and local-level edge computing guidelines need to be established to successfully handle the data generated by CVs.

Other concerns must be addressed before CV digital infrastructure will become "public ready." For example, the security of the digital infrastructure and privacy of the users must be guaranteed. Earlier studies have investigated the cyber-security aspect of V2I applications for different public infrastructure components like RSUs (Islam et al., 2017) and connected traffic signal controllers (Feng et al. 2017). USDOT has also developed the Security and Credential Management Systems to protect privacy among the communicating devices, thus ensuring security, which can be used by public agencies for digital infrastructure deployments. As shown in Figure 5, many US agencies have cyber security policies in place for ITS devices, but for V2I applications these policies will not be sufficient, as the number of external connected devices will be higher than exist today. Public agencies should collaborate with industry and academia to develop a resilient cyber-security framework for TCPS. Further, as we have noted, the standardization of these technologies across political jurisdictions must be established in order to facilitate interoperability and provide the requisite levels of both efficiency and security.

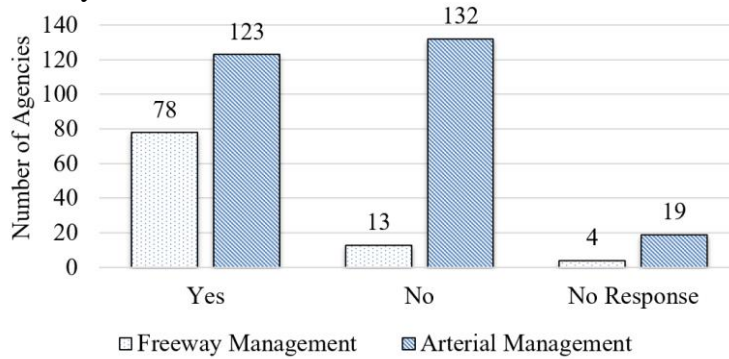

Figure 5. Agencies with Security Policies for ITS Field Devices and Communication Systems (Data from ITS Deployment Tracking 2018).

Finally, there are issues in terms of the wireless communication network. For example, as we have noted, HetNet is required for multiple V2I applications (including safety, mobility, and environmental applications). In HetNet, DSRC could provide low latency within a limited range for safety critical applications such as collision warnings, while for applications that require longer range but where higher latency is acceptable, such as queue warnings to vehicles not within a DSRC limited range, other options such as LTE may be more appropriate (Dey, et al., 2016). However, once deployed and incorporated in HetNet, the cellular 5G network could support early rollout of CV and Smart City initiatives. 5G will provide more data at a higher transmission rate for an increased number of simultaneous users in larger portions of the coverage area compared to the cellular 4G network (Rappaport et al. 2013). However, there is no mandate about specific communication options from the Federal Government for V2I applications in US, as was discussed earlier. All these technical gaps have slowed digital infrastructure investment in the transportation system.

*Investment Challenges and Opportunities*

To ensure that digital infrastructure investment is made wisely, collaboration across jurisdictions will be essential. Agencies responsible for deploying, managing and maintaining transportation infrastructure must participate in national and international platforms to plan, implement, and manage the digital infrastructure in their own jurisdictions. These plans could include national intelligent transportation plans and/or architectures, which both the US and Europe have developed. Such architectures would ensure interoperability across products, jurisdictions and regions. Failure to adopt common standards may result in haphazard deployment that would limit the efficacy of CVs.

Another strategy should involve transportation agencies following the existing CV and Smart City pilot sites' instrumentation experience (i.e., RSUs, back-end computational infrastructure, and wired/wireless communication options), moving toward deploying similar, albeit improved, technology on public roadways. Moreover, all stakeholders involved in the creation of this digital infrastructure must squarely address the remaining political, legal, and technical challenges, such as security and privacy, which have the potential to slow the development of CVs. According to a recent study, among 115 DOT personnel in Oregon, 14% thought that Oregon DOT needs to invest in legislation and regulatory actions update before allowing AVs (Bertini et al 2016).

There have been efforts at surmounting these challenges. To gain useful insights about the ITS devices' collected data, MassDOT organized "Hackathon" as a part of their Open Data initiative (MassDOT 2014). New York City followed the Open Data Government program to create a citywide open data portal where all city government agencies publish data under their jurisdiction (Dawes et al. 2016). Based on the available data, developers can create application for citizens. Also, New York City has established annual 'BigAppsNYC' competitions to encourage local entrepreneurs and researchers to develop applications. Creating such outreach programs and other similar competition/training activities will help agencies to realize the maximum benefits from V2I applications, and provide blueprints for programs that can be adopted by other cities.

Public agencies need to justify the investment in digital infrastructure to successfully implement V2I applications. At first, agencies should identify the critical areas under their jurisdiction where V2I applications will bring the maximum benefit. Based on agency requirements and application criteria, the key design considerations (i.e., computing infrastructure, communication technology) can vary. Existing literature provides sufficient data about the costs and benefits of digital infrastructure components, which can be used by public agencies to help justify funding requests (Williges et al., 2018, ITS Benefits 2018, Co-Pilot 2015).

Financing digital infrastructure investment for CVs is, of course, a major challenge. Traditional means (e.g. fuel taxes, subsidies from general funds, and tolls) may be used to generate the necessary revenue. Public-private partnerships, which have been gaining momentum worldwide, are also a promising potential mechanism to deal with upfront investment costs. Some programs are addressing this issue: for example, the American Association of State Highway and Transportation Officials (AASHTO), along with the ITS America, and Institute of Transportation Engineers (ITE), has announced the AASHTO Signal Phase and Timing (SPaT) Challenge. This encourages transportation agencies to deploy DSRC-enabled infrastructure in 20 intersections in each of the 50 states by 2020, and maintain its operations for a minimum of 10 years (Zmud et al. 2017). With this challenge, AASHTO wants not only to provide technical resources with implementation guidelines, but to identify funding sources for the participating agencies.

Whatever financing methods are chosen, the path forward for mainstreaming digital roadway infrastructure necessitates creating strong partnerships between all CV stakeholders, particularly private industry and public agencies. With foresight, commitment and cooperation, V2I may engender a unique public/private collaboration, with private industry developing the in-vehicle components of a CV system and public agencies equipping the infrastructure with complementary technology. Such a collaboration will be greatly mutually beneficial.

**CONCLUSIONS**

The benefits of CV technology are clear, multifaceted, and potentially dramatic. Travelers, automakers, insurers, public agencies, and the general public all stand to reap the rewards from an environment featuring V2I communication, including improved safety, enhanced environmental sustainability, increased mobility, and much else.

Unfortunately, the investment and regulation that are required to reach this potential is lagging. Although automakers are beginning to commit resources to CV technology for V2V communications, they are understandably reluctant to depend on public agencies given that the current public infrastructure is not ready to interact with connected vehicles. In the US, there have been federal investments, especially in the research, development, and pilot deployment of V2I infrastructure. However, investment is lacking at the state and local levels. There must be regional, national, and international collaboration to engage private and public stakeholders in planning and implementing digital infrastructure. The products of this collaboration must include platforms that will allow public and private stakeholders to cooperate in CV deployment, with common architecture and standards which would allow for interoperability between vehicles, infrastructure, and devices across political jurisdictions. This in turn will benefit private enterprise through the success of its products, as well as government entities which will be better able to provide the public improved safety and mobility across the infrastructure they operate and manage. Failure to adopt common standards may result in haphazard deployment that would limit the efficacy of CVs.

Transportation agencies should imitate the existing CV and Smart City pilot sites' instrumentation strategies, moving toward deploying similar, albeit improved, technology on public roadways. Moreover, all stakeholders involved in the creation of this digital infrastructure must squarely address the remaining political, legal, and technical challenges, such as security and privacy, which have the potential to slow the development of CVs.

With foresight, commitment and cooperation, CVs may engender a unique public/private collaboration, with private industry rapidly developing the in-vehicle and infrastructure components of a CV system, and public agencies equipping the infrastructure with such technology. Another option could be total privatization of the digital infrastructure. Such efforts, either private/public collaboration or privatization, could greatly benefit not only the stakeholders, but society as a whole.